\documentclass[aps,twocolumn,superscriptaddress]{revtex4-2}
\usepackage{graphicx}
\usepackage[colorlinks=true,linkcolor=blue,urlcolor=blue,citecolor=blue]{hyperref}
\usepackage{soul}
\usepackage{amsmath}
\usepackage{mathtools} 
\usepackage{amssymb}
\usepackage{amsthm}
\usepackage{bm,upgreek} 
\usepackage{xcolor}
\usepackage{graphicx}
\usepackage{epstopdf}

\newcommand{\nn}{{\nonumber}}
\newcommand{\bea}{\begin{eqnarray}}
\newcommand{\eea}{\end{eqnarray}}

\newcommand{\eg}{\textit{e.g.}}

\newcommand{\up}{\uparrow}
\newcommand{\dn}{\downarrow}

\newcommand{\0}[1]{\mathbf{#1}}
\newcommand{\av}[1]{\left\langle #1 \right\rangle}

\newcommand{\me}{\mathrm{e}}
\newcommand{\w}{\omega}

\begin{document}

\title{Anomalous isotope Effect in d-wave superconductors on the square lattice}

\author{Gan Sun}
\affiliation{National Laboratory of Solid State Microstructures $\&$ School of Physics, Nanjing University, Nanjing, 210093, China}
\author{Qing-Geng Yang}
\affiliation{National Laboratory of Solid State Microstructures $\&$ School of Physics, Nanjing University, Nanjing, 210093, China}
\author{Da Wang}\email{dawang@nju.edu.cn}
\affiliation{National Laboratory of Solid State Microstructures $\&$ School of Physics, Nanjing University, Nanjing, 210093, China}
\affiliation{Collaborative Innovation Center of Advanced Microstructures, Nanjing 210093, China}
\author{Qiang-Hua Wang} \email{qhwang@nju.edu.cn}
\affiliation{National Laboratory of Solid State Microstructures $\&$ School of Physics, Nanjing University, Nanjing, 210093, China}
\affiliation{Collaborative Innovation Center of Advanced Microstructures, Nanjing 210093, China}

\begin{abstract}
Isotope effect with a large coefficient $\alpha=-\partial \ln T_c/\partial \ln M$ is usually taken as an evidence of phonon mediated superconductors in the Bardeen-Cooper-Schrieffer (BCS) theory.
However, in cuprates which are now widely believed to be strong correlation induced d-wave superconductors, $\alpha$ is experimentally observed to be quite small at optimal doping, but keeps growing up with decreasing $T_c$ upon doping, even after exceeding the BCS value $1/2$.
Such an anomalous isotope effect seems to challenge the non-phonon picture and still leave room for the phonon-dominated mechanism.
In this work, we show that the anomalous dependence of $\alpha$ on $T_c$ can actually be obtained in spin fluctuation induced d-wave superconductors, by studying the Hubbard model on square lattices with functional renormalization group.
We have considered two types of electron-phonon couplings (EPCs).
The first type couples to electron densities, including the Holstein, breathing and buckling phonons, called Holstein-like.
For all these EPCs, $\alpha$ is negative and drops down towards $-\infty$ with decreasing $T_c$ upon doping.
On the opposite, for the other type of Peierls-like EPC coupling to electron hoppings on the nearest bonds, also called Su-Schrieffer-Heeger phonon, $\alpha$ is positive, grows up with decreasing $T_c$ and tends to diverge as $T_c\to0$, in qualitative agreement with the experiments.
The difference between these two types of EPCs can be understood by their isotope effects on spin fluctuations.
From this study, we conclude that the SSH phonon can explain the anomalous isotope effect in cuprates, although it is not the leading pairing mechanism.
\end{abstract}

\maketitle

\emph{Introduction}.
In the original Bardeen-Cooper-Schrieffer (BCS) theory of phonon mediated superconductivity (SC) \cite{BCS1957}, the transition temperature $T_c$ is predicted to be proportional to the Debye frequency $\w_D\propto M^{-1/2}$ with $M$ the atomic mass.
Therefore, isotope substitution with heavier $M$ reduces $T_c$, known as the isotope effect, which is described by a dimensionless isotope coefficient
\begin{eqnarray}
\alpha=-\frac{\partial \ln T_c}{\partial \ln M} = \frac12 \frac{\partial \ln T_c}{\partial \ln \w_D} .
\end{eqnarray}
Historically, the BCS value $\alpha=1/2$ had already been observed in mercury \cite{Maxwell1950,Reynolds1950} and was taken as an early evidence of the BCS theory.
Later, by taking the retardation effect \cite{Migdal1958,Eliashberg1960} and Coulomb pseudopotential \cite{Morel1962} into account, the isotope coefficient can be strongly reduced, even to negative values \cite{Carbotte1990}, but the results are still believed to be consistent with most conventional superconductors \cite{Garland1963}.

After the discovery of cuprate high-$T_c$ superconductors \cite{Bednorz1986}, the absence of isotope effect with $\alpha\approx0$ at optimal doping \cite{Batlogg1987,Bourne1987,Morris1988,Hoen1989,Yvon1989}, together with some other experiments, were immediately taken as evidences of strong correlation induced unconventional SC in cuprates \cite{Anderson1987,Zhang1988}.
However, away from the optimal doping, a series of later experiments \cite{Crawford1990, Bornemann1991, Franck1993, Soerensen1995, Zhao1996, Zhao1997, Zhao2001, Kamiya2014} have revealed a systematic behavior of the isotope effect: $\alpha$ grows up with decreasing $T_c$ upon doping away from the optimal level. More surprisingly, the growth of $\alpha$ does not stop even after it exceeds $1/2$.
Such an anomalous isotope effect clearly goes beyond the standard BCS \cite{BCS1957} and Migdal-Eliashberg theories \cite{Migdal1958,Eliashberg1960,Carbotte1990}.
Basically speaking, it tells us the electron-phonon coupling (EPC) cannot be completely neglected.
But the role of EPC, whether essential or not, on the SC mechanism is still under hot debate.
Although the strong electronic correlation is widely believed to be responsible for SC in cuprates \cite{Lee2006,Keimer2015}, the anomalous isotope effect, in particular the large $\alpha$ at low $T_c$, seems to be at odds with the non-phonon mechanism and hence leaves a large room for phonon dominated theories, such as Jahn-Teller polaronic or bipolaronic SC \cite{Alexandrov1981a,Alexandrov1981,Alexandrov1994,Alexandrov1999,Keller2005,Newns2007,Keller2008,BussmannHolder2022,Zhang2023}.
In fact, in angle-resolved photoemission spectroscopy (ARPES) experiments, two particular kinds of EPCs, the breathing and buckling phonons, are indeed found to strongly affect nodal and antinodal quasiparticles, respectively \cite{Lanzara2001,Devereaux2004}, and may also be responsible for the recently observed nearest neighbor attraction in doped cuprate chains \cite{Chen2021}.

These experimental progresses about the phonon effect in cuprates have inspired numerous theoretical studies \cite{Tsuei1990,Carbotte1991,Nazarenko1996,Pao1998,Xing1999,Dahm2000,Pringle2000,Schneider2001,Macridin2009,Macridin2012,Bechlaghem2013,Wang2015,Dzhumanov2015,Greco2016,Sun2021,Wang2021}
by combining the strong electronic correlation (beyond Coulomb pseudopotential) with different EPCs.
Most of the early studies focused on the Holstein-like EPCs, for which the phonon displacements couple to electron densities, including the Holstein, breathing and buckling phonons \cite{Song1995,Bulut1996,Pao1998,Honerkamp2007,Johnston2010}.
Recently, another type of EPC, called Peierls-like or Su-Schrieffer-Heeger (SSH) phonon \cite{SSH1979,SSH1988} coupling to electron hopping on the nearest bonds, has attracted much attention \cite{Pearson2011,Weber2015,Clay2020,Cai2021,Cai2022,Feng2022,Yang2022,Yirga2023,Zhang2023,Yang2024}.
In this work, we focus on the anomalous isotope effect and attempt to identify the roles of different EPCs in spin fluctuation induced d-wave SC (d-SC) as motivated by cuprates \cite{Tsuei2000,Scalapino2012}.
In order to treat the retarded EPC and instantaneous electronic interaction on an equal footing, we resort to functional renormalization group (FRG) \cite{Wetterich1991,Metzner2012,WangWS2012,XiangYY2012,WangWS2013,Wang2015,Yang2022}.
We find the experimental results of the isotope effect can be well reproduced by the SSH phonon but not the Holstein-like ones.
Although the SSH phonon is not the leading pairing interaction, the increase of its frequency $\w_D$ can promote spin fluctuations and hence enhances $T_c$.
In this respect, the anomalous isotope effect, in particular with $\alpha>1/2$ as $T_c\to0$, is not in contrast but in good agreement with the non-phonon SC mechanism.

\begin{figure}
\includegraphics[width=0.8\linewidth]{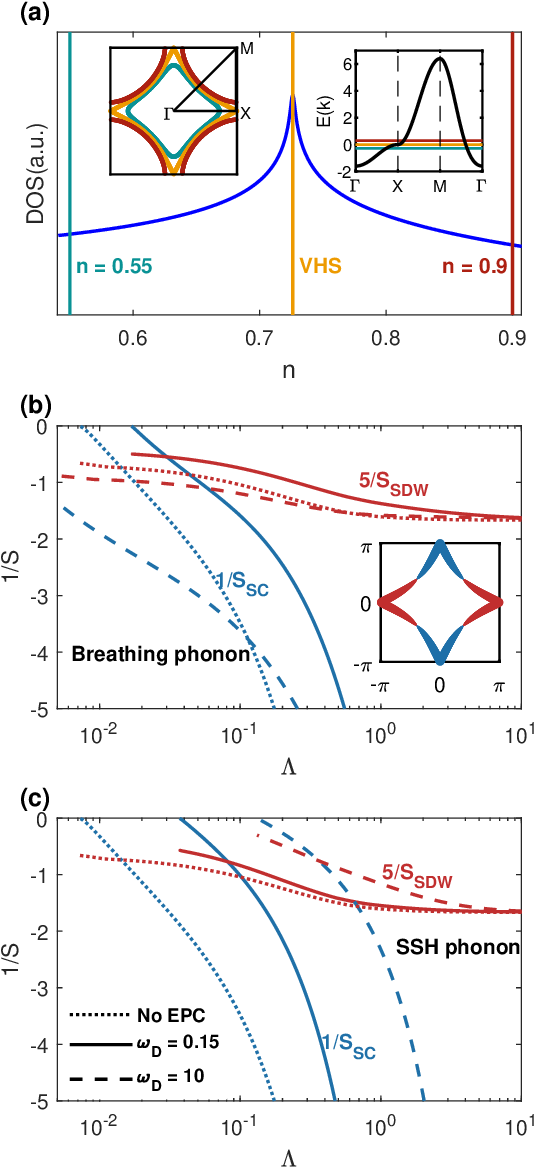}
\caption{(a) The normal state DOS versus electron filling $n$. The fermi surfaces at three fillings as indicated and the energy band dispersion are plotted in the left and right insets, respectively. (b) FRG flows of the leading $S(\0q)$ in the SDW (red) and SC (blue) channels at the VHS filling for the case of breathing phonon with $\lambda=0.02$ and $\w_D=0.15$ (solid lines), $10$ (dashed lines), in comparison with the case without EPC (dotted lines). The inset shows the gap function (sign by color and amplitude by width) on the fermi surface at divergence. (c) is similar to (b) but for the SSH phonon.}
\label{fig:flow}
\end{figure}

\begin{figure*}
\includegraphics[width=1\linewidth]{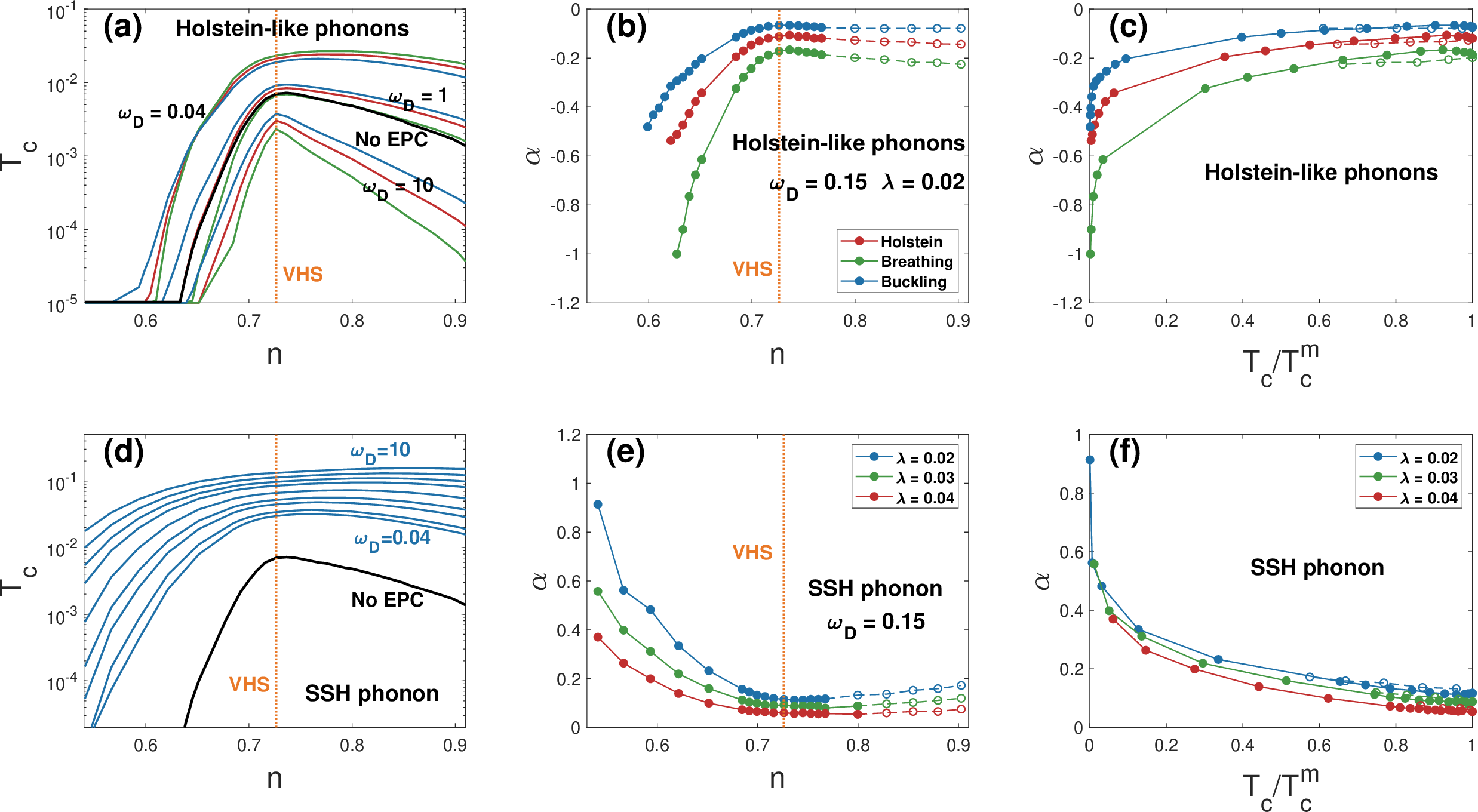}
\caption{For the three kinds of Holstein-like EPCs with $\lambda=0.02$, (a) plots $T_c$ versus $n$ for three values of $\w_D=0.04,1,10$, (b) plots $\alpha$ versus $n$ for $\w_D=0.15$, and (c) plots $\alpha$ as a function of $T_c/T_c^m$ ($T_c^m$ at optimal doping). In (b) and (c), solid and open markers denote overdoped and underdoped regimes, respectively. The results of SSH phonon are plotted in (d), (e) and (f) similarly but with three values of $\lambda=0.02$, $0.03$ and $0.04$ for comparison.}
\label{fig:isotope}
\end{figure*}

\emph{Model}.
In order to obtain the electronic correlation induced d-SC, we start from the Hubbard model on the square lattice with the Hamiltonian
\begin{eqnarray}
H=-\sum_{\av{ij}\sigma}c_{i\sigma}^\dag t_{ij} c_{j\sigma} -\sum_{i\sigma} \mu n_{i\sigma} + U\sum_{i} n_{i\up}n_{i\dn} ,
\end{eqnarray}
where $c_{i\sigma}^\dag$ creates an electron on site-$i$ with spin $\sigma$ ($\up$ or $\dn$), and $n_{i\sigma}=c_{i\sigma}^\dag c_{i\sigma}$ is the density operator.
The hopping integral $t_{ij}$ equals $t$ ($t'$) between the (next) nearest neighbor sites, $\mu$ is the chemical potential and $U$ is the Hubbard repulsion.
In this work, we choose $t$ as the energy unit, $t'=-0.3$ and $U=3$. The existence of $t'$ shifts the van Hove singularity (VHS) away from half-filling to $n_{\rm VHS}\approx0.73$ at $\mu=4t'$, as seen from the density of states (DOS) shown in Fig.~\ref{fig:flow}(a).
Near the VHS filling, the strong antiferromagnetic fluctuation can induce the d-SC \cite{Scalapino1986}.
Doping away from the VHS, the spin fluctuation is reduced and leads to weaker d-SC \cite{Metzner2012}.
Throughout this work, we confine our studies between fillings $n=0.55$ and $0.9$.
In this way, we can tune $T_c$ by doping, as shown in Fig.~\ref{fig:isotope}(a) with no EPC, serving as a starting point to further study the $T_c$-dependence of the isotope coefficient $\alpha$ for different EPCs.

In this work, we consider four kinds of EPCs. The first is the standard Holstein phonon
\begin{eqnarray}
    H'_{\rm H} = g_{\rm H} \sum_i z_i n_i,
\end{eqnarray}
which can arise from the vertical ($z$-directional) motion of the apical oxygen atom.
The second is the breathing phonon \cite{Song1995,Bulut1996},
\begin{eqnarray}
    H'_{\rm br} = g_{\rm br} \sum_i [x_{i+\frac x2} (n_i-n_{i+x}) + y_{i+\frac y2} (n_i-n_{i+y})],
\end{eqnarray}
where $x_{i+\frac x2}$ is the $x$-directional motion of the in-plane oxygen atom on the bond $(i,i+\hat{x})$. The definition of $y_{i+\frac y2}$ is similar.
The third EPC is the buckling phonon \cite{Devereaux1995,Devereaux1999,Devereaux2004,Bulut1996},
\begin{eqnarray}
    H'_{\rm bu} = g_{\rm bu} \sum_i [z_{i+\frac x2}(n_i+n_{i+x}) + z_{i+\frac y2}(n_i+n_{i+y})],
\end{eqnarray}
where $z_{i+\frac x2}$ and $z_{i+\frac y2}$ are the $z$-directional motions of the in-plane oxygen atoms {for the buckled CuO$_2$-plane}.
These three kinds of EPCs all couple to electron densities, and hence are called Holstein-like.
In addition, we also consider another kind of EPC, the SSH phonon \cite{SSH1979}
\begin{eqnarray}
    H'_{\rm SSH} = g_{\rm SSH} \sum_{\av{ij}\sigma}z_{\av{ij}}(c^{\dagger}_{i\sigma}c_{j\sigma} + H.c.),
\end{eqnarray}
where $z_{\av{ij}}$ can arise from the $z$-directional motion of the in-plane oxygen atom on each $\av{ij}$-bond {for the buckled CuO$_2$-plane}. Note that the SSH phonon couples to hopping on each bond, hence, also called Peierls-like.
For all these EPCs, for simplicity, we assume the dispersionless Einstein phonon with frequency $\w_D$ in this work.
After integrating out the phonons, we obtain the effective action $\frac12 B_{ij}^\dag \Pi_\nu B_{ij}$, where $B_{ij}$ is the fermion bilinear coupling to the phonon ($i=j$ for Holstein phonon) and $\Pi_\nu = \lambda W\omega_D^2/(\omega_D^2+\nu^2)$ is the retarded attraction with $\nu$ the Matsubara frequency, $W$ the bandwidth, and $\lambda = g^2/M\omega_D^2W$ the dimensionless EPC constant.

\emph{FRG flows}.
In order to obtain the isotope effect for the spin fluctuation induced d-SC, we need to treat pairing, spin and charge channels on an equal footing, which can be captured by FRG efficiently.
The basic idea of FRG \cite{Wetterich1991} is to obtain the one-particle irreducible (1PI) vertex $\Gamma_\Lambda$ flowing with the infrared cutoff $\Lambda$ by solving a group of FRG flow equations \cite{Metzner2012} in the presence of retarded interactions \cite{Wang2015,Yang2022}.
After obtaining the four-point $\Gamma_\Lambda$, we extract the effective interaction matrices in SC, spin-density-wave (SDW) and charge-density-wave (CDW) channels, respectively.
With decreasing $\Lambda$, the first divergence of the most negative singular value $S$ among the three channels and collective momentum $\0q$ indicates an instability with the order parameter given by the corresponding eigen scattering mode, and the divergence energy scale gives $T_c$.
More technical details of our FRG algorithm based on singular-mode decomposition \cite{WangWS2012,XiangYY2012,WangWS2013} can be found in the Supplementary Material.

Typical FRG flows of $S$, the most negative one of $S(\0q)$, versus $\Lambda$ at the VHS filling are presented in Fig.~\ref{fig:flow}(b) and (c) for the breathing and SSH phonons, respectively.
In both cases, the CDW channel is weak and not shown here.
At first, let us look at the flows without EPC.
At high energy scales, the SDW channel, as already induced by $U$, grows up with decreasing $\Lambda$.
The overlap among different channels then sets the seed for SC on bond.
At low energy scales, the SDW channel grows up slowly due to the lack of perfect fermi surface nesting.
But the SC channel always enjoys the ``perfect nesting'' between time-reversal pairs and thus grows up faster, bypasses SDW and diverges at first.
The collective momentum $\0q$ is found to be zero, indicating the Cooper instability.
From the corresponding eigen scattering mode, we find it is d-wave with the gap function projected on the fermi surface shown in the inset of Fig.~\ref{fig:flow}(b). In real space we have pairings on the nearest neighbor bonds with sign change from $x$-bond to $y$-bond.

Next, let us turn on the EPCs. In Fig.~\ref{fig:flow}(b), we plot the results of breathing phonon at $\lambda=0.02$ with $\w_D=0.15$ (solid lines) and $10$ (dashed lines).
We find the low frequency phonon with $\w_D=0.15$ can enhance the SDW interaction (relative to $\lambda=0$) at high energy scales as a result of the polaronic effect, similar to the Holstein phonon \cite{Wang2015}.
As $\Lambda$ decreases, the enhancement of SDW further increases the d-SC interaction by channel mixing and causes higher $T_c$ finally.
This is in sharp contrast to the high frequency case with $\w_D=10$, which acts as (not only) a negative-$U$ in the D-channel, hence, reduces the SDW interaction, leading to lower $T_c$.
The above analysis naturally indicates a negative isotope effect with $\alpha<0$.
In addition to the breathing phonon, we also find similar flows for the Holstein and buckling phonons (not shown).

For the SSH phonon, the flows are shown in Fig.~\ref{fig:flow}(c) with the same parameters as above.
The low frequency phonon ($\w_D=0.15$) shows similar behavior to the breathing phonon: the SDW interaction is enhanced, and hence increases $T_c$ of the d-SC (relative to $\lambda=0$).
On the other hand, the high frequency ($\w_D=10$) phonon also enhances SDW and d-SC, since the SSH phonon induced effective interaction itself has an antiferromagnetic spin-exchange component $\0S_i\cdot\0S_j$ on each nearest neighbor bond \cite{Cai2021,Yang2022}.
The latter effect is more efficient than the polaronic effect to enhance $T_c$, as seen from our FRG flows, implying a positive isotope coefficient $\alpha$.

\begin{table}
\begin{tabular}{|c||c|c|c||c|}
\hline
 & Holstein & breathing & buckling & SSH \\
\hline
$-[\Pi_\Lambda]_{\rm dSC}$ & $0$ & $+$ & $-$ & $-$ \\
\hline
$\partial V_{\rm SDW}/\partial\w_D$ & $-$ & $-$ & $-$ & $+$ \\
\hline
\end{tabular}
\caption{Signs of the effective d-SC interaction $-[\Pi_\Lambda]_{\rm dSC}$ and the isotope effect on the SDW interaction $V_{\rm SDW}$ for the four kinds of EPCs studied in this work.}
\label{table}
\end{table}

\emph{Anomalous isotope effect}.
After understanding typical FRG flows at the VHS filling, we dope the system to change $T_c$ systematically.
The results of $T_c$ versus electron filling $n$ at $\lambda=0.02$ for a series of $\w_D$ are collected in Fig.~\ref{fig:isotope}(a) for the three Holstein-like phonons.
Among them, the buckling phonon gives the highest $T_c$ because of its attractive d-SC interaction $-[\Pi_\Lambda]_{\rm dSC}$, as listed in Table~\ref{table}.
From the $\w_D$-dependence (not explicitly shown), the isotope coefficient $\alpha$ is obtained and plotted versus $n$ at $\w_D=0.15$ in Fig.~\ref{fig:isotope}(b).
Finally, from the doping-dependence of $T_c$ and $\alpha$, we extract $\alpha$ as a function of $T_c/T_c^m$ ($T_c^m$ at optimal doping) directly, as shown in Fig.~\ref{fig:isotope}(c).
The three kinds of Holstein-like phonons show similar behaviors: $\alpha$ is always negative and its amplitude $|\alpha|$ grows up with decreasing $T_c$ upon doping.
Most astonishingly, $\alpha$ seems to diverge towards $-\infty$ as $T_c\to0$.

On the other hand, for the SSH phonon, the results are shown in Fig.~\ref{fig:isotope}(d,e,f) in the same way as above.
The behaviors of $\alpha$ are quite different from the Holstein-like ones.
We always obtain positive $\alpha$.
It grows up with decreasing $T_c$ upon doping and also tends to diverge but towards $+\infty$ as $T_c\to0$.
For comparison, we also show the results of $\lambda=0.03$ and $0.04$.
We find larger $\lambda$ gives higher $T_c$ (due to its attractive d-SC interaction) but smaller $\alpha$, leading to similar plots of $\alpha$ versus $T_c/T_c^m$.
Here, the $T_c$-dependence of $\alpha$ for the SSH phonon, is in qualitative agreement with the experiments in cuprates.

The above results for the Holstein-like and SSH phonons are clearly beyond the BCS and Migdal-Eliashberg theories \cite{Carbotte1990,Sun2021} in which only the SC channel is considered.
Therefore, the key to understand these results should be the mutual effect among different channels, as we have captured in our FRG.
For the Holstein-like EPCs, higher frequency phonon reduces the SDW interaction, which in turn suppresses $T_c$ of the d-SC.
In the opposite, for the SSH phonon, higher frequency one enhances the SDW interaction and then increases $T_c$.
To describe these effects more quantitatively, we can build a phenomenological model to describe the channel mixing, by simply allowing the dimensionless pairing interaction $\lambda_{\rm SC}$ (not to confuse with EPC constant $\lambda$) depending on the phonon frequency $\w_D$.
As a comparison, in BCS and Migdal-Eliashberg theories, $\lambda_{\rm SC}$ and $\w_D$ are independent quantities.
But here, since the total pairing interaction mainly comes from (non-phonon) spin fluctuations, $\partial\lambda_{\rm SC}/\partial\w_D$ can be nonzero, as listed in Table~\ref{table}.
In this way, even from the standard BCS $T_c$-formula $T_c=1.13\w_D\me^{-1/\lambda_{\rm SC}}$, we obtain the isotope coefficient
\begin{eqnarray}
\alpha=\frac12 + \frac12 \frac{\partial\lambda_{\rm SC}}{\partial\ln\w_D} \ln^2\left(\frac{T_c}{1.13\w_D}\right) .
\end{eqnarray}
Clearly, if $\lambda_{\rm SC}$ is independent of $\w_D$, we immediately get back to the BCS value $1/2$.
But if $\partial\lambda_{\rm SC}/\partial\w_D\ne0$, we always obtain divergent $\alpha\sim\pm\ln^2T_c$ as $T_c\to0$, where $\pm$ depends on the sign of $\partial\lambda_{\rm SC}/\partial\w_D$.
This simply explains what we obtained in FRG.

Interestingly, we have also obtained this scaling behavior of $\alpha\sim\pm\ln^2T_c$ as $T_c\to0$ in BCS and Eliashberg theories with two boson modes \cite{Sun2021}.
Together with the wide observation of growing up of $\alpha$ with decreasing $T_c$ in both cuprates and Sr$_2$RuO$_4$ \cite{Mao2001}, we suggest that $\alpha\sim\pm\ln^2T_c$ as $T_c\to0$ may be a property for many superconductors.
Actually, it seems the only exception is the standard BCS superconductor with $\alpha\equiv1/2$.
Back to cuprates, although our studies are limited to moderate interaction not yet in the realm of Mott limit, our results still have strong implications:
The observation of increasing $\alpha$ to values above $1/2$ with decreasing $T_c$ is not inconsistent with the non-phonon mechanism. Instead, it may be an evidence that the phonon is not the leading pairing interaction. This reconciles the longstanding paradox between the strong correlation SC mechanism and the anomalous isotope effect in cuprates.

\emph{Summary}.
We studied the isotope effect of Holstein-like (Holstein, breathing, buckling) and Peierls-like (SSH) EPCs on $T_c$ of the spin-fluctuation induced d-SC.
With decreasing $T_c$ upon doping away from the optimal level, $\alpha$ is negative and drops down for the Holstein-like phonons, but is positive and grows up for the SSH phonon. The latter is consistent with the experiments in cuprates.
For both types of EPCs, $\alpha$ diverges to $\pm\infty$ as $T_c\to0$, which can be understood by their different isotope effects on spin-fluctuations.
Our discovery suggests the anomalous isotope effect in cuprates does not support, at the face value, a phonon-dominated mechanism for SC. Instead, it merely unravels the nontrivial effect of electron-phonon coupling on the spin fluctuations that eventually lead to d-SC.

\emph{Acknowledgement}.
This work is supported by National Key R\&D Program of China (Grant No. 2022YFA1403201) and National Natural Science Foundation of China (Grant No. 12374147, No. 12274205, No. 92365203, and No. 11874205).
The numerical calculations were performed at the High Performance Computing Center of Nanjing University.

\bibliography{isotope}

\begin{widetext}

\begin{center}
\Large Supplementary Materials
\end{center}

This supplementary material is divided into two parts.
In the first part, we introduce the singular-mode functional renormalization group (SM-FRG) method in the presence of phonon mediated retarded interactions. Some technical details are included.
In the second part, we supplement additional results for comparison including: the FRG flows for the Holstein and buckling phonons, and two cases (circular Fermi surface and $U=0$) with spin fluctuations strongly suppressed.
\section{Singular-mode functional renormalization group}

Functional renormalization group (FRG) is a method to study strongly interacting system based on the exact flow equations of some generating functions.
General introductions to FRG can be found in many good reviews (see \eg \cite{Metzner2012} and references therein).
In this Supplementary Material, we will only focus on our specific implementation, called singular-mode FRG (SM-FRG) \cite{WangWS2012,XiangYY2012,WangWS2013}, by applying it to interacting fermionc systems with phonon mediated retarded interactions \cite{Wang2015,Yang2022,Yang2024}.

Our implementation is based on the Wetterich scheme \cite{Wetterich1991} by choosing the effective action $\Gamma$ as the generating function, which then generates the FRG flows of the $m$-point one-particle irreducible (1PI) vertices $\Gamma^{(m)}$ with respect to the infrared cutoff $\Lambda$. However, in most applications, it is impossible and not necessary to keep all orders of $\Gamma^{(m)}$, and three simplifications can be made:
(1) By power counting, all $m$-point vertices $\Gamma^{(m)}$ with $m>4$ are irrelevant and thus can be safely neglected supposing the instability occurs at sufficiently low energy scale.
(2) For 4-point vertices, only the instantaneous (frequency-independent) terms are marginal, while the frequency-dependent ones are irrelevant. Therefore, we only need to keep the flows of the instantaneous 4-point vertices. (Phonon mediated retarded interactions are taken into account by successively adding into the instantaneous ones, see Sec.~\ref{sec:epc} for implementation.)
(3) In addition, the 2-point vertex (self-energy) is also neglected since its effect can be absorbed into the band structure if we are working near the Fermi liquid fixed point.

\subsection{Three Mandelstam channels and flow equation}

\begin{figure}
\includegraphics[width=0.5\textwidth]{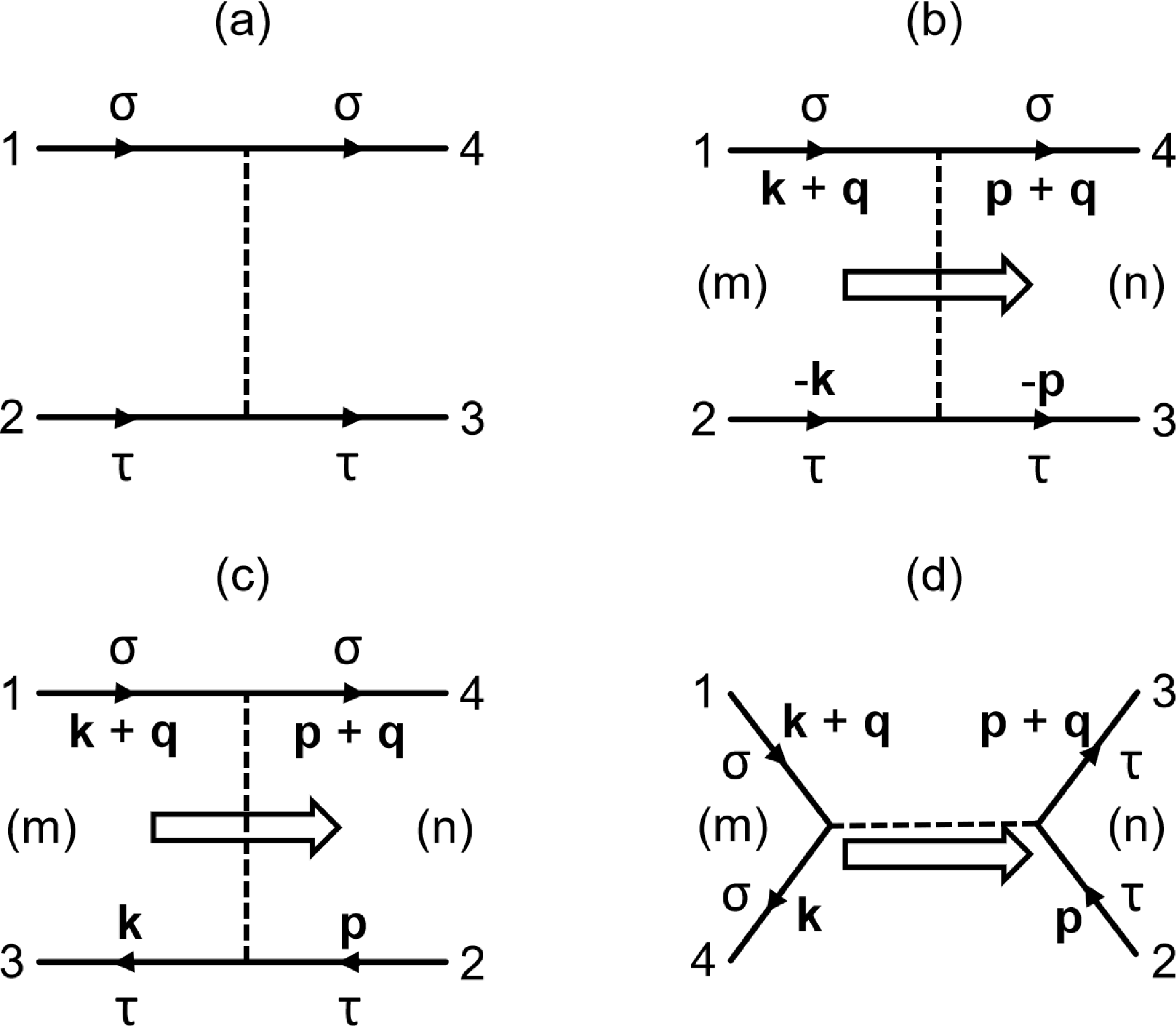}
\caption{A general 4-point 1PI vertex (a) is rewritten into the pairing ($P$), crossing ($C$) and direct ($D$) channels, as shown in (b), (c) and (d), respectively. In these plots, $k$, $p$ and $q$ are 4-momenta, $\sigma$ and $\tau$ denote spins, $(m)$ and $(n)$ indicate form factors of the fermion bilinears.}
\label{fig:gamma4_PCD}
\end{figure}

As mentioned above, we are interested in the FRG flows of the 4-point instantaneous 1PI vertices $\Gamma_{1234}$ (superscript ``$(4)$'' omitted henceforth), as shown in Fig.~\ref{fig:gamma4_PCD}(a), where the subscripts $1\sim4$ denote one-particle degrees of freedom. For the spin SU(2) symmetric case, spins $\sigma$ and $\tau$ are conserved during fermion propagation.
Our SM-FRG is based on the efficient parameterization of $\Gamma_{1234}$ in three Madelstam channels specified by fermion bilinears as,
\begin{align}
\alpha_{12}^{\dag}=\psi_1^\dag\psi_2^\dag \ \text{ (pairing)}, \quad \beta_{13}^{\dag}=\psi_1^\dag\psi_3\  \text{ (crossing)}, \quad \gamma_{14}^{\dag}=\psi_1^\dag\psi_4 \ \text{ (direct)}.
\end{align}
Then the 4-point vertex can be rearranged as scattering matrices $P$, $C$ and $D$ in the three channels as illustrated in Fig.~\ref{fig:gamma4_PCD}(b)-(d),
\begin{align}
\frac{1}{2}\sum_{1,2,3,4}\psi^\dagger_1\psi^\dagger_2\Gamma_{1234}\psi_3\psi_4&=\frac12 \sum_{12,43}\alpha_{12}^{\dag} ~P_{12;43}~\alpha_{43}=\frac12\sum_{qmn}\alpha_{qm}^\dag P_{mn}(q) \alpha_{qn} \\
&= -\frac12 \sum_{13,42}\beta_{13}^{\dag} ~C_{13;42} ~\beta_{42}=-\frac12\sum_{qmn}\beta_{qm}^\dag P_{mn}(q) \beta_{qn} \\
&= \frac12 \sum_{14,32}\gamma_{14}^{\dag} ~D_{14;32} ~\gamma_{32} \label{eq:rewind}=\frac12\sum_{qmn}\gamma_{qm}^\dag P_{mn}(q) \gamma_{qn}
\end{align}
where $q=k_1+k_2$, $k_1-k_3$, $k_1-k_4$ is the collective 4-momentum of the two fermion bilinears in the $P$, $C$ and $D$ channels, respectively, and the remaining degrees of freedom are grouped into bilinear labels $m$ and $n$.
If the subscripts $1,2,3,4$ run over all sites, the representations with three scattering matrices $P$, $C$ and $D$ are all equivalent to $\Gamma$, i.e. $\Gamma_{1234}=P_{12;43} =C_{13;42}=D_{14;32}$.
But in practical calculations, the fermion bilinears must be truncated. On physical grounds, the important bilinears are those that join the singular scattering modes, and such eigen modes determine the emerging order parameters. Since the order parameters are composed of short-ranged bilinears, only such bilinears are important. These include onsite and on-bond pairings in the $P$ channel, and onsite and on-bond particle-hole densities in the $C$ and $D$ channels. The FRG based on the decomposition of the 1PI vertices into scattering matrices in the truncated fermion bilinear basis, which are sufficient to capture the most singular scattering modes, is called the singular-mode FRG (SM-FRG) \cite{WangWS2012,XiangYY2012,WangWS2013}.

\begin{figure}
\includegraphics[width=0.55\linewidth]{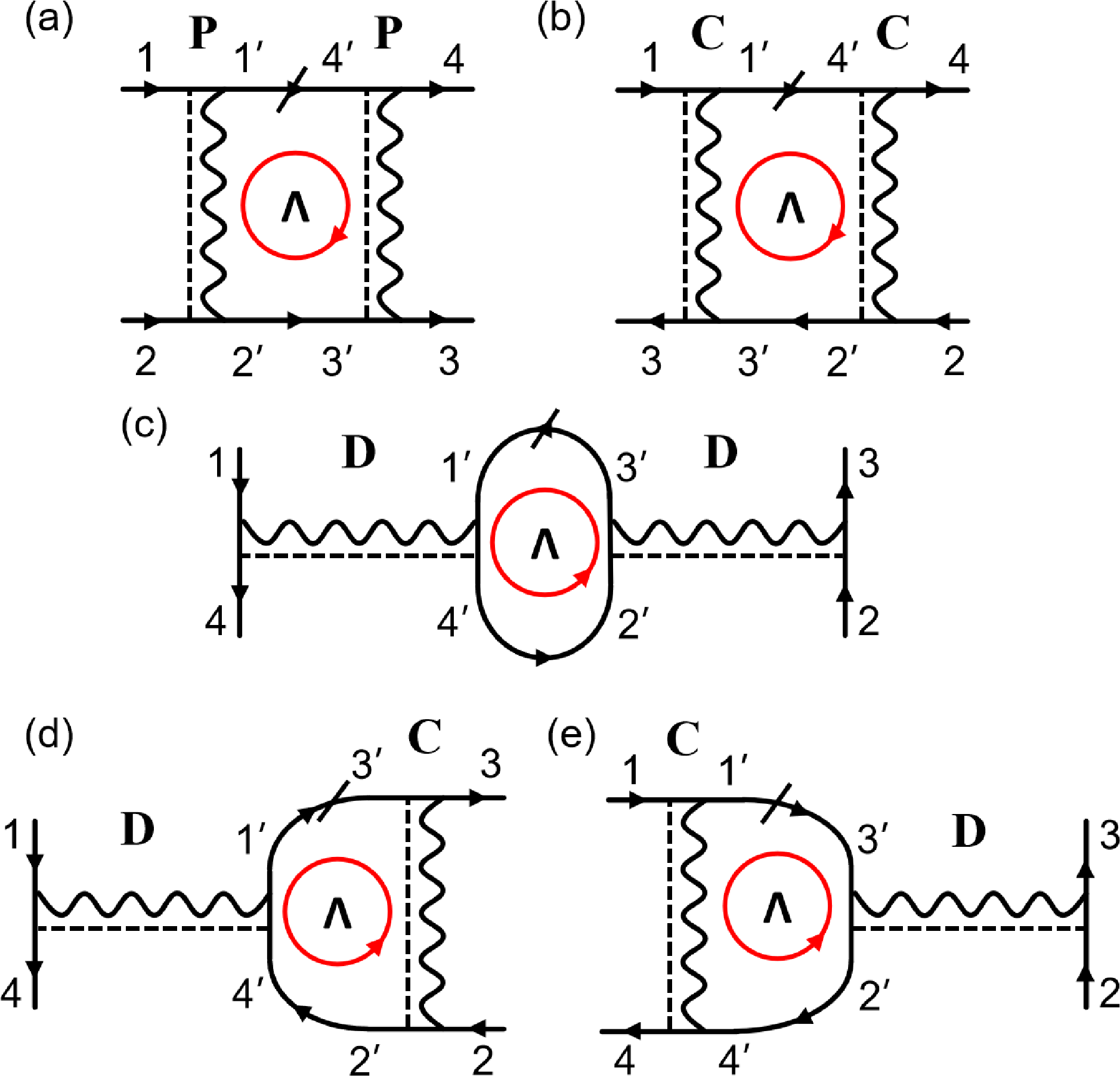}
\caption{Contributions to $\partial\Gamma_{1234}/\partial\Lambda$.
The dashed line denotes the instantaneous vertex ($P$, $C$, or $D$) and the wavy line denotes the bare phonon meditated retarded interaction $\Pi_\nu$ projected in the respective channel. They are added up in the calculation. The slash denotes the single-scale propagator and can be put on either one of the fermion lines within the loop. The frequency along the directed circle is $\Lambda$ as indicated.}
\label{fig:gamma4_1loop}
\end{figure}

Starting from $\Lambda=\infty$ where the 1PI vertices $P$, $C$ and $D$ are given by the bare interactions, $\Gamma_{1234}$ follows the differential flow equation,
\begin{align} \label{eq:flow}
\frac{\partial\Gamma_{1234}}{\partial\Lambda} = [P\chi_{pp}P]_{12;43}+[C\chi_{ph}C]_{13;42} + [D\chi_{ph}C+C\chi_{ph}D-2D\chi_{ph}D]_{14;32} .
\end{align}
The five terms on the right hand side are illustrated in Fig.~\ref{fig:gamma4_1loop}.
The products within the square brackets should be understood as convolutions, and $\chi_{pp}$ and $\chi_{ph}$ are single-scale (at $\Lambda$) particle-particle and particle-hole susceptibilities given by, in real space,
\begin{align}
	[\chi_{pp}]_{ab;cd} &= \frac{1}{2\pi}\left[ G_{ac}(i\Lambda)G_{bd}(-i\Lambda)+
	(\Lambda \to -\Lambda) \right], \\
	[\chi_{ph}]_{ab;cd} &= \frac{1}{2\pi}\left[ G_{ac}(i\Lambda)G_{db}(i\Lambda)+
	(\Lambda \to -\Lambda) \right],
\end{align}
where $a,b,c,d$ are dummy fermion indices (that enter the fermion bilinear labels), and $G_{ab}(i\Lambda)$ is the normal state Green's function. The expression in momentum space is straightforward, and is used in actual calculations.
The functional flow equation Eq.~\ref{eq:flow} is solved by numerical integration over $\Lambda$.
Note that after $\Gamma_{1234}$ is updated during each integration step, it is rewinded as $P$, $C$ and $D$ according to $\Gamma_{1234}=P_{12;43} =C_{13;42}=D_{14;32}$, subject to truncation of the fermion bilinears to be discussed below.
In this way, SM-FRG can treat interactions in all channels on equal footing. In fact, if we ignore the channel overlaps, the flow equation reduces to the ladder approximation in the $P$ channel, and the random phase approximation in the $C$ and $D$ channels. The SM-FRG combines the three channels coherently.

\subsection{Phonon mdediated retarded interaction}
\label{sec:epc}

For electron-phonon coupled systems, the phonon degrees of freedom can be integrated out exactly to obtain a retarded electron-electron interaction $\Pi_\nu$, if we neglect the phonon-phonon interactions. But as mentioned above, the frequency-dependent 4-point 1PI vertices are RG irrelevant. Therefore, we only need to trace the flows of the instantaneous (frequency-independent) vertices and add the effect of $\Pi_\nu$ step by step during the FRG flow.
The contributions to the flows of $\Gamma$ are illustrated in Fig.~\ref{fig:gamma4_1loop}, where the dashed lines denote instantaneous vertices, and the wavy lines are from the retarded kernel suitably added to the instantaneous part. Explicitly, the flow equation can be written as
\begin{align} \label{eq:ph-flow}
\frac{\partial\Gamma_{1234}}{\partial\Lambda} = [\+P\chi_{pp}\+P]_{12;43}+[\+C\chi_{ph}\+C]_{13;42}
+[-2\+D\chi_{ph}\+D + \+D\chi_{ph}\+C + \+C\chi_{ph}\+D]_{14;32} ,
\end{align}
where $\+P=P+[\Pi_\Lambda]_P$, $\+C=C+[\Pi_\Lambda]_C$, $\+D=D+[\Pi_0]_D$, with $[\Pi_\nu]_{P,C,D}$ the projection of $\Pi_\nu$ in the respective channels (or associated to the desired fermion bilinears).
Note that the frequency of $\Pi_\nu$ is $\Lambda$ in $\+P$, $\+C$, and $0$ in $\+D$, as a consequence of the frequency conservation since the external fermions can be freely set at zero frequency for instantaneous vertices.

\subsection{Effective interactions and order parameters}

\begin{figure}
\includegraphics[width=0.75\linewidth]{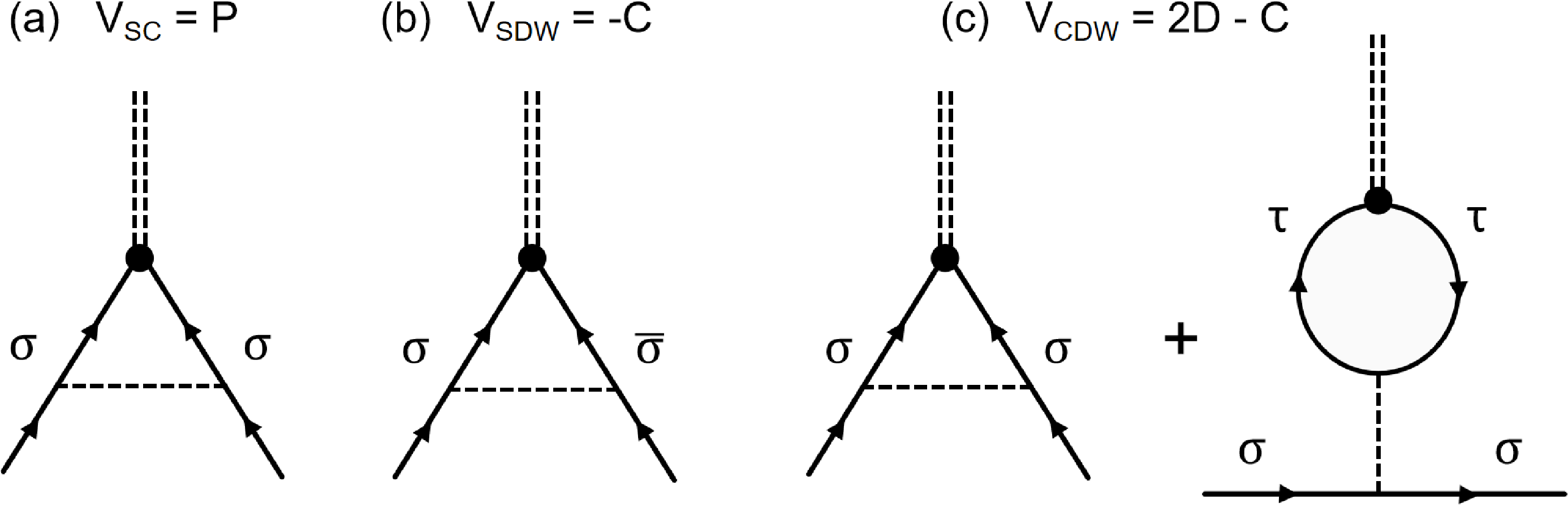}
\caption{By analyzing the 3-point vertices coupling to different external fields (double dashed lines), we obtain the effective interactions in the SC (a), SDW (b), and CDW (c) channels, respectively. $\sigma$ and $\tau$ denote spin, and $\bar{\sigma}$ denotes $-\sigma$.}
\label{fig:V}
\end{figure}

From $P$, $C$, and $D$, we can extract the scattering matrices, or effective interactions, in the superconductivity (SC), spin-density-wave (SDW) and charge-density-wave (CDW) channels, respectively,
\begin{align}
    V_{\rm SC}=P,\ \ V_{\rm SDW}=-C,\ \ V_{\rm CDW}=2D-C.
\end{align}
This can be understood by considering the 3-point vertices coupling to external fields in the three channels, respectively, as illustrated in Fig.~\ref{fig:V}.

In a given channel, the effective interaction matrix can be decomposed by singular value decomposition (SVD), in momentum space, as
\begin{align}
V_{mn}(\0q)=\sum_{\alpha} \phi_{\alpha m}(\0q) S_{\alpha}(\0q) \phi^*_{\alpha n}(\0q),
\end{align}
where $m,n$ label the fermion bilinear, $S_\alpha$ and $\phi_{\alpha n}$ are eigen value and vector for the $\alpha$-th singular mode.
During the SM-FRG flow, we monitor the leading (most negative) eigenvalue, which we abbreviate as $S$ for simplicity in each channel.
As the energy scale $\Lambda$ reduces, the first divergence of $S$ indicates a tendency towards an instability with the order parameter described by the associated eigen mode $\phi(\0Q)$, where $\0Q$ is the associated collective momentum.
In this case, one can drop the nonsingular components to write the renormalized interaction as,
\begin{equation}
H_{\Gamma} \sim \frac SN O^\dagger O+\cdot\cdot\cdot,
\end{equation}
where $N$ is the number of unitcells, $O$ is the mode operator that is a combination of the fermion bilinears (see below), and the dots represent symmetry related terms.
For example, if the SC channel diverges first, we have
\begin{equation}
O^\dagger_{\rm SC} = \sum_n \phi_n(\0Q) \alpha_n^\dagger(\0Q) \to \sum_{\0k,n=(a,b,\bm{\delta})}\psi_{\0k+\0Q, a}^\dagger \phi_n(\0Q) e^{i\0k\cdot\bm{\delta}}\psi_{-\0k,b}^\dagger,
\end{equation}
where $\alpha^\dagger_{ij}=c^\dagger_ic^\dagger_j$ represents the fermion bilinear in the $P$ channel, $n$ labels a fermion bilinear, $a$ and $b$ denote other fermion degrees of freedom (such as orbital and sublattice), and $\Delta_{\bm{\delta}}^{ab}=\phi_n(\0Q)$ is the element of the real-space pairing matrix on the bond $\bm{\delta}$ radiating from $a$.
The spin indices do not have to be specified, as the symmetry of the gap function under inversion automatically determines whether the pair is in the singlet or triplet state.

Similarly, if the SDW channel diverges first, we obtain the mode operator
\begin{equation}
O^\dagger_{\rm SDW} = \sum_n \phi_n(\0Q)\beta_n^\dagger(\0Q)\to \frac{1}{\sqrt{N}}
\sum_{\0k,n=(a,b,\bm{\delta})}\psi_{\0k+\0Q,a,\uparrow}^\dagger \phi_n(\0Q) e^{i\0k\cdot\bm{\delta}}\psi_{\0k,b,\downarrow},\end{equation}
where $\beta^\dagger_{ij}=c^\dagger_ic_j$ represents the fermion bilinear in the $C$ channel, and the spin order is assigned in the transverse direction.

Finally, if the CDW channel diverges first, we obtain the mode operator
\begin{equation}
O^\dagger_{\rm CDW} = \sum_n \phi_n(\0Q)\gamma_n^\dagger(\0Q)\to \frac{1}{\sqrt{N}}
\sum_{\0k,\sigma,n=(a,b,\bm{\delta})}\psi_{\0k+\0Q,a,\sigma}^\dagger \phi_n(\0Q) e^{i\0k\cdot\bm{\delta}}\psi_{\0k,b,\sigma},\end{equation}
where $\gamma^\dagger_{ij}=c^\dagger_ic_j$ represents the fermion bilinear in the $D$ channel.
Note that $H_{\rm SDW/CDW}$ can capture both onsite and on-bond density waves, since the fermion bilinears contain both cases of $\bm{\delta} = \00$ and $\bm{\delta} \neq \00$.

\subsection{Truncation of fermion bilinears}

\begin{figure}
\includegraphics[width=0.45\linewidth]{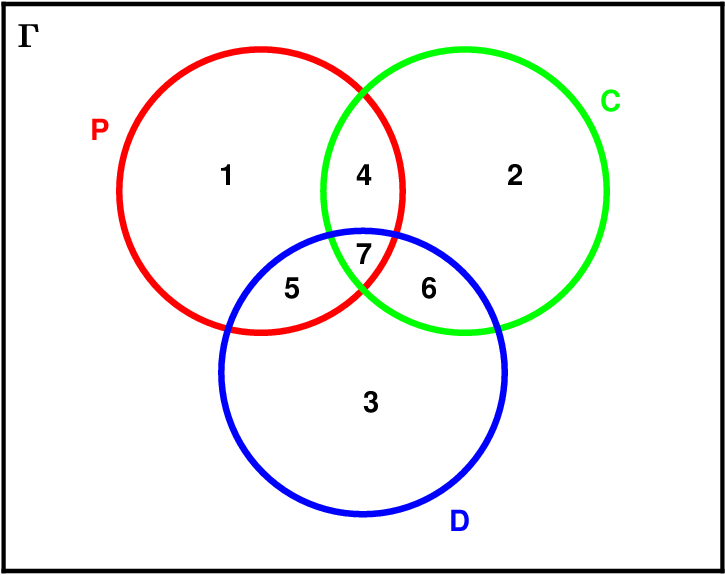}
\caption{Schematic diagram illustrating the relationship between the truncated 1PI vertices $P$, $C$ and $D$. }
\label{fig:G_PCD}
\end{figure}

In practical calculations, it is impossible and not necessary to retain all possible fermion bilinears. Only short-ranged bilinears are sufficient to capture the most singular scattering modes as discussed above.
Therefore, we perform truncation in real space by imposing a truncation length $L_c$ on the internal relative distance between two fermions within each bilinear as
\begin{align}
|\0R_2-\0R_1|\leq L_c, \ \ |\0R_3-\0R_4|\leq L_c, \ \ \text{for }P, \nn\\
|\0R_3-\0R_1|\leq L_c, \ \ |\0R_2-\0R_4|\leq L_c, \ \ \text{for }C, \nn\\
|\0R_4-\0R_1|\leq L_c, \ \ |\0R_2-\0R_3|\leq L_c, \ \ \text{for }D.
\end{align}
Due to the truncation, the truncated 1PI vertices $P$, $C$ and $D$ are no longer equivalent to $\Gamma_{1234}$, and consequently there are partial overlaps between the three, as shown in Fig.~\ref{fig:G_PCD}.
For example, if $|\0R_2-\0R_1|\leq L_c$, $|\0R_3-\0R_1|\leq L_c$, $|\0R_3-\0R_4|\leq L_c$ and $|\0R_2-\0R_4|\leq L_c$ are simultaneously satisfied, then $P_{12;43}=C_{13;42}$, which corresponds to the region of $1\cap2=4\cup7$ in Fig.~\ref{fig:G_PCD}.

\begin{figure}[h]
\includegraphics[width=0.6\linewidth]{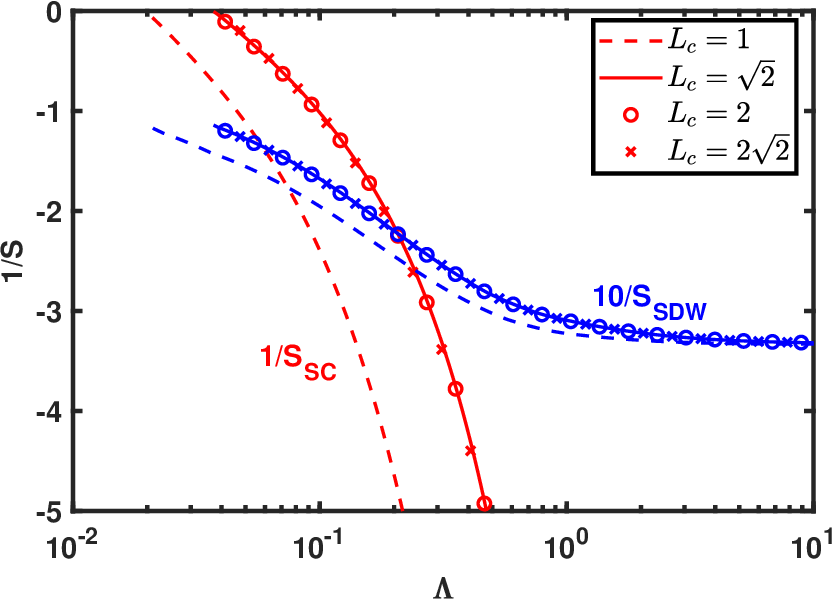}
\caption{FRG flows of the leading $S(\0q)$ in SDW (blue) and SC (red) channels for the case of SSH phonon at the van Hove filling by choosing different truncation lengths. In the calculations, $U=3$, $\w_D=0.15$ and $\lambda=0.02$.
}
\label{fig:Lc}
\end{figure}

In the main text, we select a truncation length $L_c=\sqrt{2}$ for the two-dimensional square lattice.
This choice allows us to include up to the next-nearest neighboring bonds in the fermion bilinear base. (Here, we set the lattice constant to $1$.)
In Fig.~\ref{fig:Lc}, we present our SM-FRG results at $L_c=1, \sqrt{2}, 2, 2\sqrt{2}$ with $U=3$, $\w_D=0.15$, $\lambda=0.02$ at the van Hove filling.
The results of $L_c=\sqrt{2}$, $2$, and $2\sqrt{2}$ almost merge together, indicating the truncation of $L_c=\sqrt{2}$ is large enough to be a very good approximation for this work.

\subsection{Self-adaptive momentum points}

\begin{figure}
\includegraphics[width=0.7\linewidth]{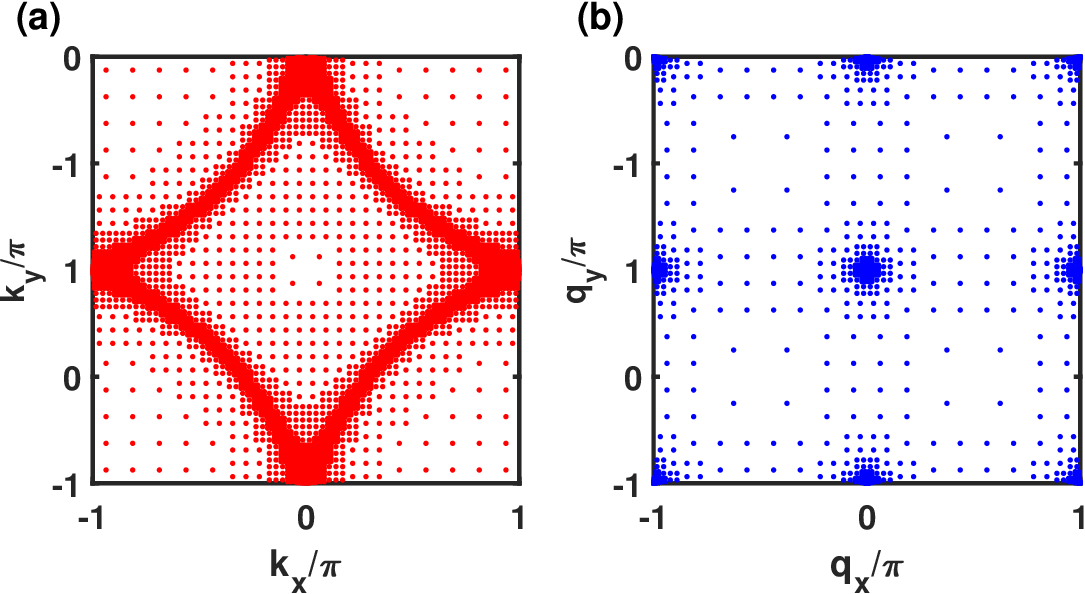}
\caption{Self-adaptive $\0k$-mesh (a) and $\0q$-mesh (b) at the van Hove filling.}
\label{fig:kmesh_qmesh}
\end{figure}

In our SM-FRG, the 4-point 1PI vertices are parametrized and stored in $P_{mn}(\0q)$, $C_{mn}(\0q)$, $D_{mn}(\0q)$, where $m$, $n$ are indices of the truncated fermion bilinears, and $\0q$ is the propagating momentum of the bilinear (as a boson). At each integration step in solving the flow equation, we perform inverse Fourier transformation to real space such that the overlaps among $P$, $C$, $D$ can be treated faithfully. After that, we perform Fourier transformation again to convert these vertices back to $\0q$-space. Such a cyclic procedure demands a high resolution in the $\0q$-space, in particular near $(0,0)$ (forward scattering) and $(\pi,\pi)$ (connecting van Hove singularities) in the present work, since the leading scattering processes mainly occur near these momenta. In practice, we design a self-adaptive procedure to generate the $\0q$-mesh, as shown in Fig.~\ref{fig:kmesh_qmesh}(b), which is progressively denser near the momenta $(0,0)$ and $(\pi,\pi)$ as desired.

On the other hand, in order to calculate $\chi_{pp}^\Lambda(\0q)$ and $\chi_{ph}^\Lambda(\0q)$, one also needs to do integration over the fermion momentum $\0k$, which then requires a high resolution of $\0k$ near the Fermi surface. For this purpose, we generate the self-adaptive $\0k$-mesh as shown in Fig.~\ref{fig:kmesh_qmesh}(a), which is progressively denser in approaching the Fermi surface.

\section{More results}

\subsection{FRG flows for Holstein and buckling phonons}

\begin{figure}
\includegraphics[width=0.65\linewidth]{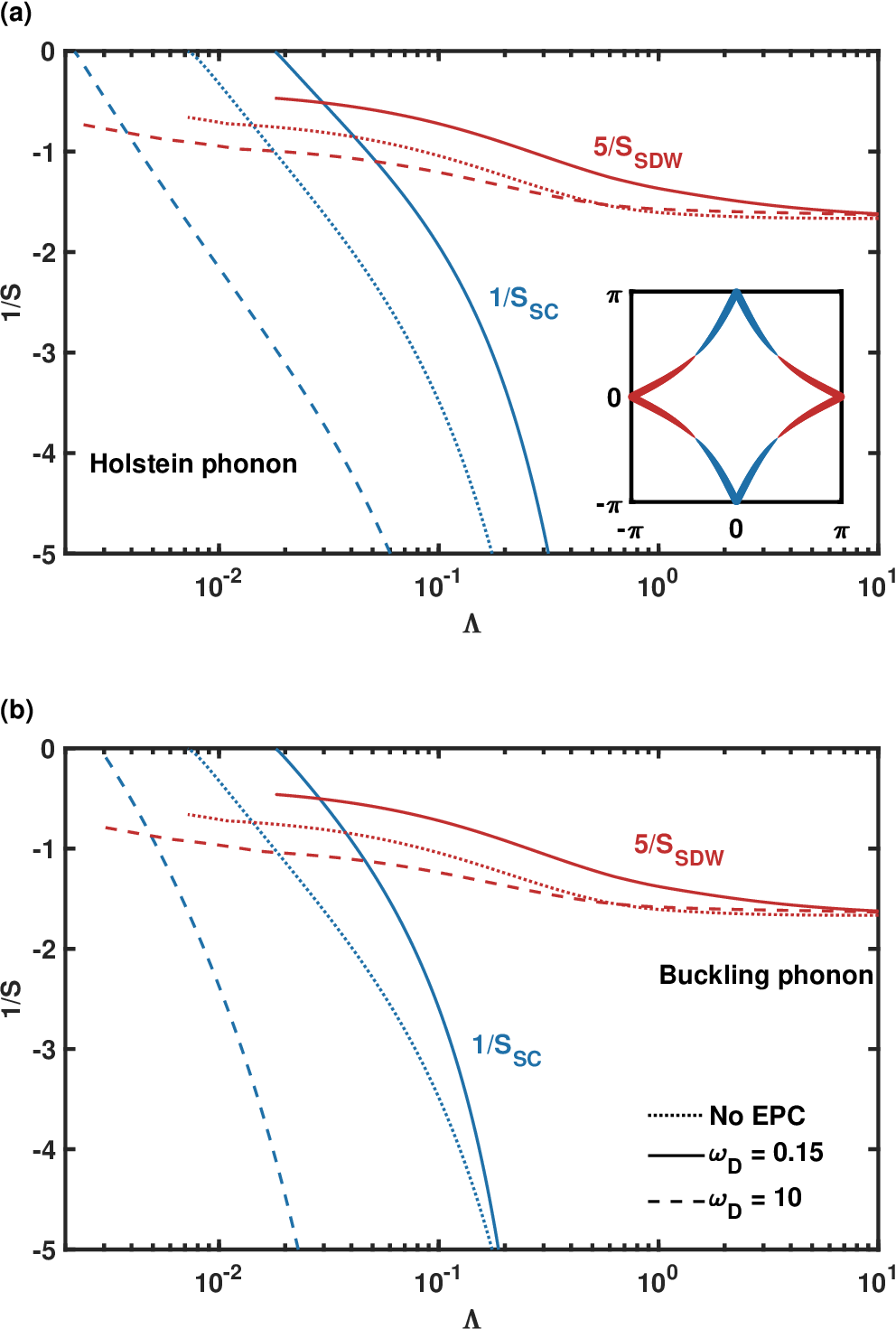}
\caption{(a) FRG flows of the leading $S(\mathbf{q})$ in the SDW (red) and SC (blue) channels at the van Hove filling for the case of Holstein phonon with $\lambda=0.02$ and $\omega_D=0.15$ (solid lines), $10$ (dashed lines), in comparison with the case without EPC (dotted lines). The inset shows the SC gap function (sign by color and amplitude by width) on the fermi surface at divergence. (b) is similar to (a) but for the buckling phonon. }
\label{fig:sflow_Holstein_Buckling}
\end{figure}

In the main text, we claim that the FRG flows of the Holstein and buckling \cite{Devereaux1995,Devereaux1999,Devereaux2004,Bulut1996} phonons are similar to the breathing phonon \cite{Song1995,Bulut1996}. In this Supplementary Material, we provide their typical FRG flows in Fig.~\ref{fig:sflow_Holstein_Buckling}(a) and (b), respectively, at $\lambda=0.02$ with $\w_D=0.15$ (solid lines) and $10$ (dashed lines). The results without EPC (dotted lines) are also given for comparison. Clearly, these results are similar to those of the breathing phonon as shown in Fig.~1(b) and 1(c) in the main text.

\subsection{Suppression of spin fluctuations for circular Fermi surface}

In this work, we find the spin fluctuation induced d-wave SC can exhibit positively diverging $\alpha$ for $T_c\to0$. As comparisons, we consider another two cases without (strong) spin fluctuations: low filling fraction yielding a circular Fermi surface in this subsection, and $U=0$ in the next subsection.

\begin{figure}
\includegraphics[width=0.6\linewidth]{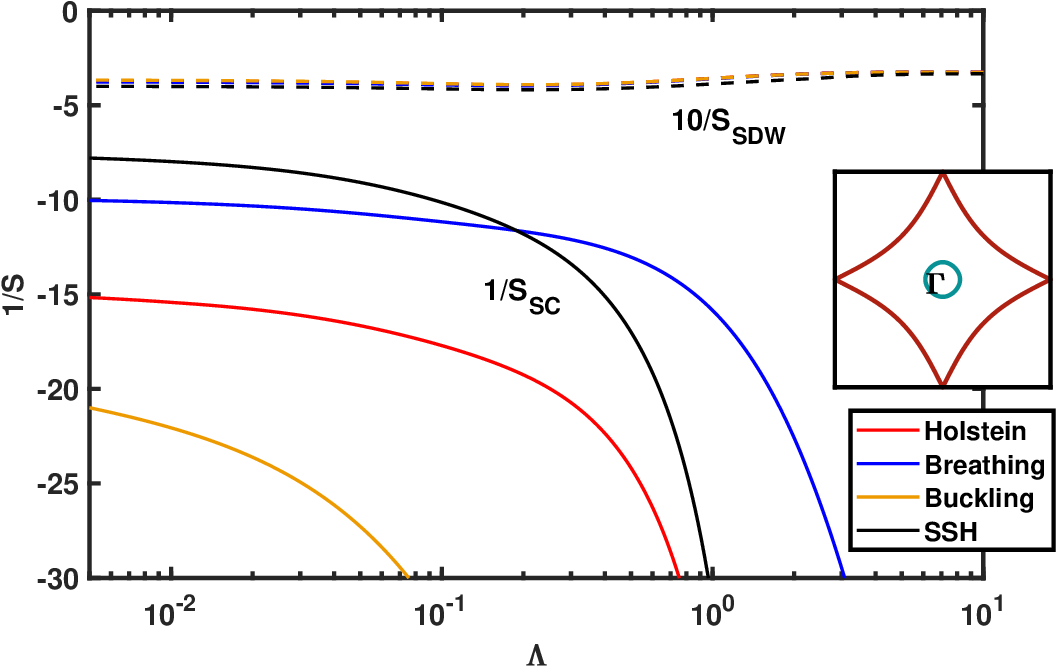}
\caption{The FRG flows of the leading $S(\0q)$ in the SC (solid line) and SDW (dashed line) channels at low filling $n=0.04$ to achieve circular Fermi surface as shown in the inset. In the calculations, we set $\lambda=0.02$ and $\omega_D=0.15$ for the four types of EPCs.}
\label{fig:smaller_n}
\end{figure}

By adjusting the chemical potential $\mu$ to filling level $n=0.04$, we obtain a circular Fermi surface around the Brillouin zone center, as shown in the inset of Fig.~\ref{fig:smaller_n}. For this case with the featureless Fermi surface, the FRG flows are shown in the main panel of Fig.~\ref{fig:smaller_n}.
The SDW channel is almost scale-independent, due to the lack of featured particle-hole scattering.
As a result, SC can only be driven by EPC. For the weak EPC strength $\lambda=0.02$ here, we observe no SC instability within the numerical range of $\Lambda>10^{-4}$.
Of course, one can enhance $\lambda$ to obtain SC, which actually comes back to the BCS-Eliashberg results with $\alpha\leq0.5$ \cite{Carbotte1990}, consistent with the following subsection.

\subsection{Suppression of spin fluctuations for $U=0$}

\begin{figure}
\includegraphics[width=0.9\linewidth]{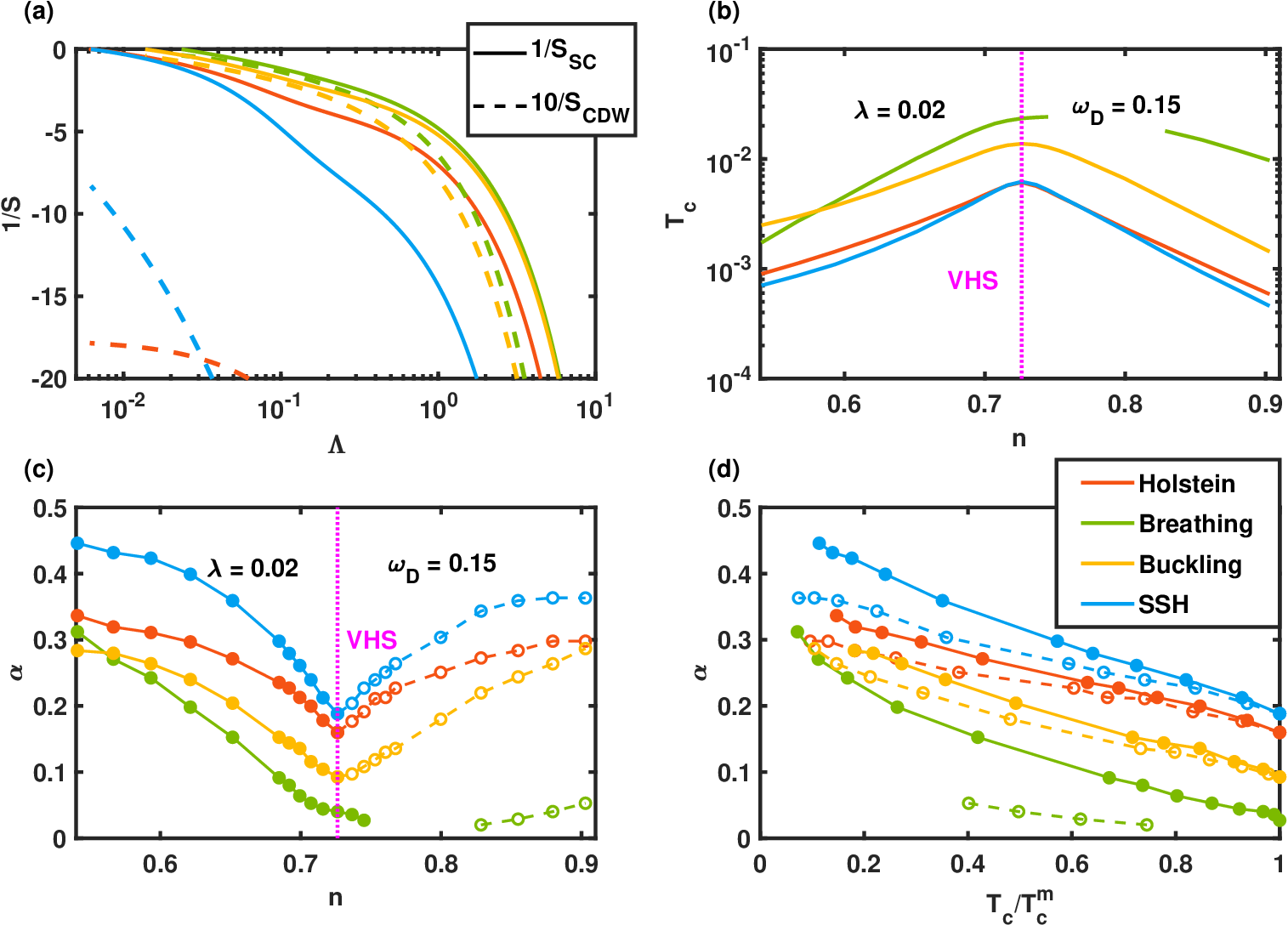}
\caption{ Results of $U=0$. (a) FRG flows of the leading $S(\mathbf{q})$ in the SC (solid line) and CDW (dotted line) channels at the van Hove filling. (b) and (c) plot doping dependence of $T_c$ and $\alpha$, respectively. (d) plots $\alpha$ versus $T_c/T_c^m$ ($T_c^m$ at optimal doping). For all the four types of EPCs, we set $\lambda=0.02$ and $\omega_D=0.3$.}
\label{fig:U0}
\end{figure}

Now we consider another way to suppress the spin fluctuations by turning off the Hubbard $U$.
The FRG flows at the van Hove filling are shown in Fig.~\ref{fig:U0}(a). For $U=0$, SDW becomes the weakest and hence not shown. Instead, the CDW channel dominates at high energy scales and induces the s-wave SC eventually.
Next, we tune $T_c$ by doping. The doping dependence of $T_c$ and $\alpha$ are presented in Fig.~\ref{fig:U0}(b) and \ref{fig:U0}(c), respectively. (For the breathing phonon, CDW instability occurs firstly in a small region.)
We find common behavior for the four types of EPCs. At the van Hove filling, $T_c$ becomes the highest due to the strong CDW fluctuations, while $\alpha$ is found to be the smallest, reflecting the negative isotope effect of CDW. Upon doping away from the van Hove filling, SC is gradually suppressed by weaker CDW fluctuations. Meanwhile, $\alpha$ grows up and approaches the BCS-Eliashberg value $0.5$ (without Coulomb pseudopotential). The dependence of $\alpha$ on $T_c$ is plotted directly in Fig.~\ref{fig:U0}(d), to compare with Fig.~2(c) and 2(f) in the main text.
These results confirm the important role of spin fluctuations in explaining the positively diverging $\alpha$ as $T_c\to0$.

\end{widetext}

\end{document}